\documentclass[twocolumn,a4paper,prl]{revtex4}
\usepackage{amsmath,amssymb,graphicx}

\begin{document}
\title{Variation of Orbital Symmetry of the Localized $3d^1$ Electron
of the V$^{4+}$ Ion upon the Metal-Insulator Transition in VO$_2$}

\author{ M. A. Korotin, N.A. Skorikov, and V. I. Anisimov}

\affiliation{Institute of Metal Physics, 620219 Ekaterinburg GSP-170,
Russia}

\begin{abstract} The electronic structure of the vanadium dioxide VO$_2$ in the
tetragonal $R$ and two monocinic $M_1$ and $M_2$ structural modifications was
calculated in frames of the local-density functional (LDA) approach and the
LSDA + U formalism of correction for correlation effects. Based on the results
of calculation, we argue in favor of the Mott-Hubbard mechanism of the
metal-insulator transition. It is shown that the transition is accompanied by
the change of the orbital in which the $3d^1$ electron of the V$^{4+}$ ion is
localized. The symmetry of the filled orbital is determined by the local oxygen
environment of the vanadium atoms in each structural phase. Depending on the
dispersion of the band that corresponds to the orbital occupied, the
investigated compound can be either a metal or an insulator. 
\end{abstract}
\maketitle

The metal-insulator transition in VO$_2$ is observed at a temperature of 340~K
and is accompanied by a structural transition from the nonmagnetic
high-temperature tetragonal rutile phase $R$ to the nonmagnetic monoclinic
$M_1$ phase. The rutile phase contains chains of equidistant vanadium atoms
along the [001] direction. The unit cell of the $M_1$ phase doubles as compared
to rutile, all vanadium atoms slightly deviate from the [001]$_R$ direction,
forming zigzags and become nonequidistant, forming pairs. 

A noncontradictory theoretical description of the metal-insulator transition in
vanadium dioxide is still lacking. The origin of the nonmetallic state is
explained now from two viewpoints. 

If one starts from the results of band calculations in the local-density
approximation (LDA), then the $M_1$ phase can be considered as a usual Peierls
band insulator \cite{Goodenough71,Wentzcovitch94}. From the other viewpoint, it
is impossible to ignore electron correlations in the $d$ shell of vanadium and
they cannot be considered on the average, as is done in band calculations. A
quarter of century ago, one more insulating phase ($M_2$) with local magnetic
moments at some vanadium atoms was discovered. The $M_2$ phase originates upon
slight doping of VO$_2$ with chromium ($\gtrsim$ 0.l~at.~\%)
\cite{crystaldata-M2} or upon application of a small stress along the [110]$_R$
axis \cite{Pouget75}. In the $M_2$ phase, in one chain, vanadium atoms form
pairs along the [110]$_R$ direction, as in the $M_1$ phase (but without
deviation from the [110]$_R$ direction), while in the neighboring chain the
atoms of vanadium again become equidistant, as in the $R$ phase, but the zigzag
configuration is preserved, as in the $M_1$ phase \footnote{The diagram of
mutual arrangement of V atoms in the three VO$_2$ phases can be found, for
example, in [\cite{review98}, Fig. 215]. The phase diagram is shown in
[\cite{review98}, Fig. 217]. The triclinic $T$ phase of VO$_2$ will not be
considered in this work.}. Because of the existence of a chain of equidistant
V$^{4+}$ ions, the band-theoretical description of the insulating state of the
$M_2$ phase becomes impossible: these chains are magnetic (Mott-Hubbard)
insulators. Therefore, theoretical models were developed in which the $M_1$ and
$M_2$ phases were considered as the Mott-Hubbard insulators, excluding the
interatomic Hubbard correlations in the $R$ phase~\cite{Zylbersztejn75}. At the
same time, arguments and experimental evidences indicating that all structural
modifications of VO$_2$ must be considered as Mott-Hubbard rather than as band
insulators appeared in the literature
\cite{Lederer72,Pouget75,Pouget74,Shin90,Nikolaev92,Rice94,review98}.

Discussion of the nature of the insulating state in VO$_2$ continues with
allowance for new experimental and theoretical data on the electronic structure
of the $R$ and $M_1$ phases. From comparison of the results of spectroscopic
measurements and band calculations, it was concluded in \cite{Kurmaev98} that
the electronic structure of VO$_2$ is well described within the local-density
approximation without an account of any correlation effects. Experimental
measurements of the temperature dependence of the velocity of surface acoustic
waves in the VO$_2$ single crystal near the metal-insulator transition
permitted Maurer \textit{et al.} ~\cite{Maurer99} to conclude that it is
unlikely that the driving force of the metal-insulator transition are electron
correlations. Measurements of the delay time in the VO$_2$ metal/oxide/metal
structures and the investigation of the influence of the injection of carriers
on switching and metal-insulator transition in planar switching devices
\cite{Stefanovich00} also confirm exclusively the Mott-Hubbard mechanism of
transition in VO$_2$. It is known that the usual band calculations indicate
only a tendency for the formation of an energy gap in the $M_1$ phase of VO$_2$
\cite{Wentzcovitch94} and give a resulting metallic spectrum in contrast to the
experiment. An attempt to resolve this contradiction within the band approach
was made. Thus, the introduction of the electron-phonon interaction into the
combination of the three-dimensional periodic-shell model and the X$\alpha$
cluster method of discrete variation resulted in the appearance of a gap in the
band spectrum \cite{Nakatsugawa97}. The introduction of the correction for
self-energy within the model $GW$ scheme also satisfactorily reproduces the
properties of the non metallic $M_1$ phase of  VO$_2$ observed experimentally
\cite{Continenza99}. Note that no theoretical calculations of the electronic
structure of the insulating $M_2$ phase are available at present.

In this work, we tried to analyze in detail the electronic structure of the
$R$, $M_1$ and $M_2$ phases of VO$_2$ (for the first time in the case of the
$M_2$ phase), in each case paying attention to the orbital that is filled by
the sole electron of the V$^{4+}$ ion. With this aim in mind, we used the
first-principles linearized MT-orbital method of calculation in the
tight-binding approximation (TB LMTO)~\cite{Andersen84,Andersen86}. The
nonmagnetic calculations were carried out within the conventional formalism of
the local-density functional (LDA). In magnetic calculations, the electron
correlations were taken into account within the LSDA+U
approach~\cite{lda+u,lda+u_modern}. We used the experimentally determined
values of the lattice parameters available in the literature (from
\cite{crystaldata-R} for the $R$ phase, from \cite{crystaldata-M1} for the
$M_1$ phase, and from \cite{crystaldata-M2} for the $M_2$ phase). In
calculations, the radii of the atomic spheres of V and O were chosen to be
equal with a 16\% overlap; empty spheres were introduced to fill the volumes of
the atomic spheres to the volume of the unit cell; the basic functions were
$4s$, $4p$, and $3d$ for vanadium, $3s$, $2p$, and $3d$ for oxygen, and $1s$,
$2p$, and $3d$ for the empty spheres. Within the LSDA formalism, we calculated
the Coulomb $U$ and exchange $J$ interactions in the $d$ shell of vanadium. The
value of $U$ was calculated with allowance for screening of the $t_{2g}$
electrons by the $e_g$ electrons~\cite{calc_U}. The values $U = 3.8$~eV and $J
= 1.0$~eV obtained were used unchanged in calculations of the electronic
structure of all phases\footnote{The existing experimental estimates of the
Coulomb interaction $U$ in VO$_2$ ~\cite{Shin90} give 2.1~eV for the insulating phase and
1.3~eV for the metallic phase. These values are slightly below that used in
calculations $U = 3.8$~eV.}. Thus, our calculations do not contain adjustable
parameters and, in this sense, are the first-principles.

\begin{figure}[t]  
\centerline{ \includegraphics [clip=true,width=53mm]{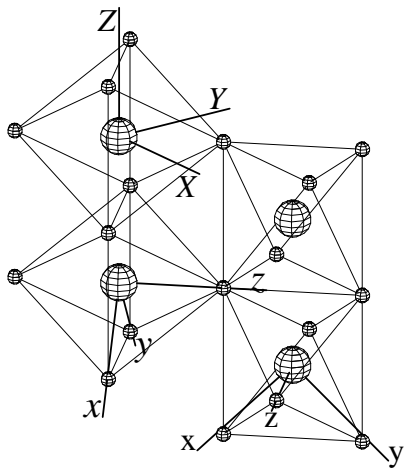} }  
\caption{The crystal structure of the $R$ phase of  VO$_2$. Large spheres are
the vanadium atoms, small spheres are the oxygen atoms. The axes of the
crystallographic coordinate system are labeled by capital letters. The axes of
the local systems of coordinates for the corner and body-center vanadium atoms are
labeled by small letters.}  
\label{R.crstr}  
\end{figure}

The unit cell of the high-temperature $R$ phase of VO$_2$ (space group
$P4_2/mnm$) contains two formula units. The vanadium atoms form a body-centered
tetragonal lattice and are surrounded by slightly deformed oxygen octahedra. In
such octahedra, the V--O distances in the direction normal to the tetragonal
$c$ axis are slightly greater than the V--O distances in the plane parallel to
the $c$ axis (Fig. \ref{R.crstr} and table). In the same plane, the O-V-O
angles are 84.2$^\circ$ and 95.8$^\circ$, while the other O-V-O angles are
equal to 90$^\circ$. The octahedra surrounding the vanadium atoms in the
vertices and in the center of the unit cell are rotated by 90$^\circ$ about the
$c$ axis. In what follows, we will present the results for each vanadium atom
in the local coordinate system, in which the $z$ axis is directed toward one of
the most remote oxygen atoms (apical atom) in the octahedra. The direction of
the chain of vanadium atoms along the $c$ axis is the bisector of the angle
which is made by the $x$ and $y$ axes of the local coordinate system
(see Fig.~\ref{R.crstr}). The Euler angles used for passage from the
crystallographic coordinate system to the local one for the vanadium atom at
the position $(0; 0; 0)$ are the set $(\frac \pi 4; \frac \pi 2; -\frac \pi
4)$.

\begin{table}[bp]
\caption{The V--O distances (\AA) for the $R$ and $M_1$ phases and the average
distances for the $M_2$ phase of VO$_2$. The "apical" and "in-plane" oxygen
atoms are considered in the local coordinate systems: the $z$ axis passes
through the apical oxygen and the $x$ and $y$ axes pass through the in-plane
oxygen atoms.}
\center
\begin{tabular} {|l|c|c|c|c|}
\hline
\multicolumn {1} {|l|} {}&
\multicolumn {1} {c|} {}&
\multicolumn {1} {c|} {}&
\multicolumn {2} {c|} {$M_2$}\\ 
\cline{4-5}
\multicolumn {1} {|c|} {Oxygen atoms}&
\multicolumn {1} {c|} {$R$}&
\multicolumn {1} {c|} {$M_1$}&
\multicolumn {1} {c|} {in the }&
\multicolumn {1} {c|} {in the }\\
\multicolumn {1} {|c|} {}&
\multicolumn {1} {c|} {}&
\multicolumn {1} {c|} {}&
\multicolumn {1} {c|} { dimerized}&
\multicolumn {1} {c|} { undimerized }\\
\multicolumn {1} {|c|} {}&
\multicolumn {1} {c|} {}&
\multicolumn {1} {c|} {}&
\multicolumn {1} {c|} { chains}&
\multicolumn {1} {c|} { chains}\\
\hline
Apical & 1.93 & 1.91 & 1.87 & 1.93 \\
In-plane & 1.92 & 1.94 & 1.97 & 1.95 \\\hline
\end{tabular}
\label{V-O.dist}
\end{table}

Let us clarify the terms that will be used in the interpretation of the
results. By occupancy of the $i$th orbital, we mean the integral $N_i = \int
_{-\infty} ^{\varepsilon_F} n_i(\varepsilon) d\varepsilon$ where
$n_i(\varepsilon)$ is the corresponding partial density of states, and
$\varepsilon_F$ is the Fermi energy. By the center of gravity of the $i$th
band, we mean the ratio 
$\varepsilon^{cg}_i \equiv <\varepsilon_i> = \int_{\varepsilon_1}
^{\varepsilon_2} \varepsilon \cdot n_i(\varepsilon)d\varepsilon \diagup
\int_{\varepsilon_1} ^{\varepsilon_2} n_i(\varepsilon) d\varepsilon$,
where  $\varepsilon_1$ and $\varepsilon_2$ are the boundaries
of the energy interval under consideration. By the dispersion of the $i$th
band, the expression  $\sigma_i = \sqrt{ <\varepsilon_i^2> -
<\varepsilon_i>^2}$ is meant. Because of the distortion of the oxygen
octahedra, the $t_{2g}$ orbitals of the vanadium atom become nondegenerate.
The $xy$, $yz-zx$, and $yz+zx$ wave functions, where $x$, $y$, and $z$
represent the basis of the local system of coordinates, will be the natural
basis for the $t_{2g}$ orbitals.

\begin{figure}[t]
\centerline{ \includegraphics [clip=true,width=65mm]{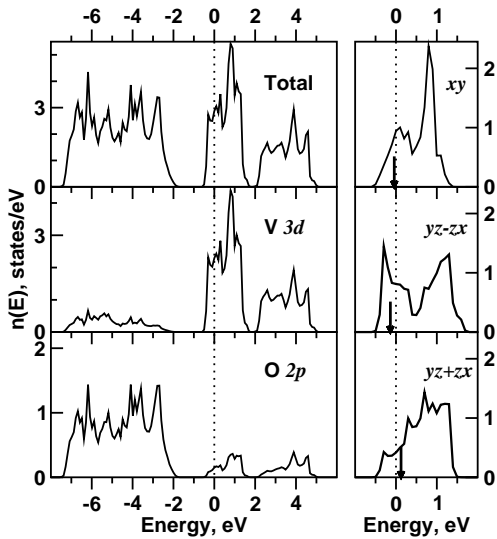} }  
\caption{The total (per formula unit) and partial (per atom) densities of
states for the $R$ phase of VO$_2$ (left column) and the $t_{2g}$ orbital
partial density of states for the vanadium atoms (right column). The positions
of the Fermi level are shown by dotted lines. Arrows in the right column
indicate the centers of gravity of the orbital partial densities of states in
the energy range from --8 to +6~eV.} 
\label{R.dos} 
\end{figure}

The results of the nonmagnetic LDA calculation are shown in Fig.~\ref{R.dos}.
Our LDA calculations show: (1) owing to the elongation of the oxygen octahedra
along the local $z$ axis, the $yz-zx$ orbital is most populated; (2) the center
of gravity of the $yz-zx$ band lies most deeply in the energy range from --8 to
+6~eV; (3) in the energy interval from --1.5 to +1.8~eV, the dispersion of the
$yz-zx$ band is approximately one and half times greater than that for the $xy$
and $yz+zx$ bands. 

\begin{figure}
\centerline{ \includegraphics [clip=true,width=40mm]{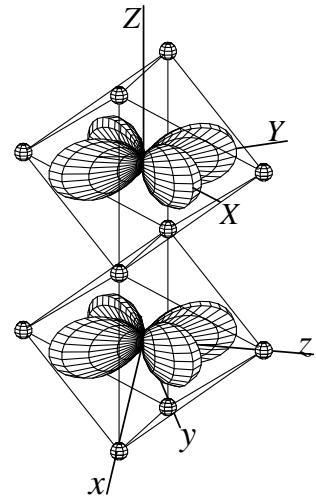} } 
\caption{Ordering of the occupied $yz-zx$ orbitals in the chain of the
vanadium atoms obtained as a result of the LSDA+U calculation for the $R$ phase
of VO$_2$ For designations, see Fig.~\ref{R.crstr}. The orbitals shown represent
the angular distribution of the $d$ electron density of the vanadium atoms 
$\rho(\theta,\phi) = \sum_{m,m'} Q_{m,m'} Y_m(\theta,\phi)
Y_{m'}(\theta,\phi)$, where $Q_{m,m'} = n^{\uparrow}_{m,m'} -
n^{\downarrow}_{m,m'}$ is the difference between the population matrices of the
$d$ states of the vanadium atoms with spins up and down obtained by
self-consistent calculation, and $Y_m(\theta,\phi)$ are the corresponding
spherical harmonics.}
\label{R.orb}
\end{figure}

We also carried out LSDA+U calculations for the $R$ phase of VO$_2$ for the
ferromagnetic and antiferromagnetic types of ordering of magnetic moments of
vanadium ions in the chains located along the tetragonal $c$ axis. In the first
case, the half-metallic ferromagnetic state is obtained with the spin magnetic
moment of the $d$ shell of vanadium equal to $\mu$=0.99~$\mu_B$. In the second
case the antiferromagnetic state was found to be insulating with an energy gap
of 0.38~eV and $\mu$=0.90~$\mu_B$. The second solution was lower than the first
one in the total energy. For both solutions the sole $d$ electron of the
V$^{4+}$ ion was localized in the $yz-zx$ orbital (Fig. \ref{R.orb}). This
naturally follows from the above results ((1) and (2)) of the LDA calculations:
within the LSDA+U approach, the orbital that already was most populated and
whose center of gravity was already the lowest among all the $t_{2g}$ orbitals
is occupied. The third result (the maximum dispersion of the $yz-zx$ band) has
preserved the ferromagnetic state as the metallic one even within the LSDA+U
calculation and led to the appearance of the antiferromagnetic insulating state
with a small energy gap\footnote{For $U \le 3.0$~eV, the antiferromagnetic
state of the $R$ phase in the LSDA+U calculation is metallic. See also [29].}. 

The fact that the LSDA+U calculations predict the antiferromagnetic insulating
state for the $R$ phase of VO$_2$ instead of the metallic one reflects the
known tendency of the LSDA+U method to overestimate the electron localization
as a result of consideration of the on-site spin- and orbit-dependent Coulomb
repulsion in terms of the mean-field theory. At the same time, the LSDA+U
approach gives an accurate description of the orbital nature of the localized
electrons\footnote{This is shown in the series of our works devoted to the
investigation of orbital filling in CrO$_2$~\cite{Korotin98} KCuF$_3$
perovskites ~\cite{lda+u_modern}, PrMnO$_3$~\cite{Anisimov97}, layered
vanadates of the CaV$_n$O$_{2n+1}$-type~\cite{Korotin99}, etc.}. Our
calculations of the electron structure of VO$_2$ in the $R$ phase predict
localization of the $d$ electron of the V$^{4+}$ ion on the $yz-zx$ orbital. As
will be seen below, this result is a key one for the description of the
metal-insulator transition in VO$_2$.

\begin{figure}[t]
\centerline{ \includegraphics [clip=true,width=65mm]{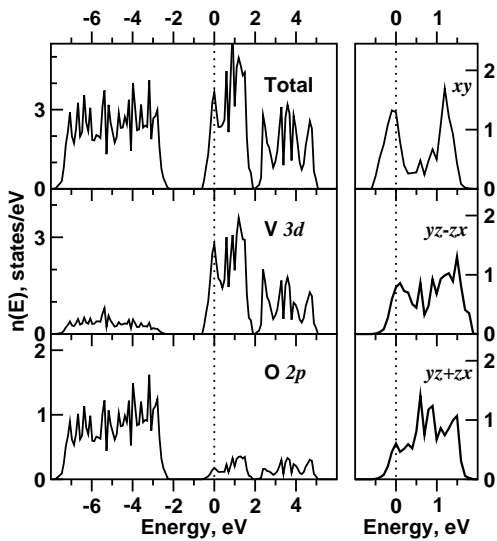} }  
\caption{The total (per formula unit) and partial (per atom) densities of
states for the $M_1$ phase of VO$_2$ (left column) and the $t_{2g}$ partial
densities of states for the vanadium atoms (right column). The Fermi level is
shown by dotted lines.}  
\label{M1.dos} 
\end{figure}

\begin{figure}[t]
\centerline{ \includegraphics [clip=true,width=75mm]{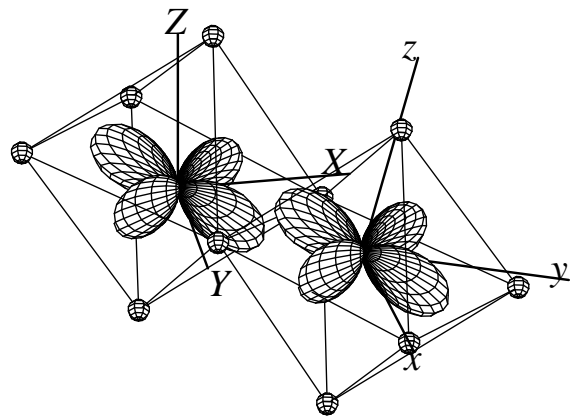} } 
\caption{Ordering of the occupied $xy$ orbitals in the chain of vanadium
atoms obtained as a result of the LDA+U calculation for the $M_1$ phase of
VO$_2$. The axes of the cristallographyc system of coordinates are marked by
capital letters. The axes of the local system of coordinates for the vanadium
atoms are marked by small letters.} 
\label{M1.orb} 
\end{figure}

Analysis of the partial densities of the $t_{2g}$ states of vanadium, obtained
in the nonmagnetic LDA calculation for the $M_1$ phase and shown
in Fig.~\ref{M1.dos}, indicates a different sequence of filling orbitals, as
compared to the $R$ phase. The $M_1$ phase (the space group $P2_1/c$) is
characterized by a pairing of the vanadium atoms along the $[0.842; 0; -0.539]$
vector in the crystallographic coordinate system: there are short and long
vanadium-vanadium bonds. Moreover, each pair of vanadium atoms is inclined
relative to this vector in the plane with the normal $[0.381; 0.708; 0.595]$
for one chain of vanadium atoms and to the plane with the normal $[0.381;
-0.708; 0.595]$ for the neighboring chain. The difference in the oxygen
surroundings of the vanadium atoms is not only in the variation of the O-V-O
angles, but also in that the V-O distance to the apical oxygen atoms in the
$M_1$ phase became smaller than to the in-plane oxygens (see
table). We define the local system of coordinates for the $M_1$
phase in the following way: the $z$ axis is perpendicular to the plane of the
vanadium atoms that belong to the same chain, and the $x$ and $y$ axes are
deflected by an angle of 45$^\circ$ from the direction of the chain
(Fig.~\ref{M1.orb}). In other words, the local system of coordinates for the
$M_1$ phase is chosen similar to that for the $R$ phase. Then, for the vanadium
atom in the $(0.259; -0.025; -0.136)$ position, the Euler angles for passing
from the crystallographic system of coordinates to the local one will be
$(0.658 \pi; -0.297 \pi; -0.984 \pi)$. 

Comparing the partial densities of the $t_{2g}$ states of vanadium for the $R$
and $M_1$ phases (see Figs.~\ref{R.dos} and \ref{M1.dos}), note a substantial
difference in the density of states for the $xy$ orbital. Owing to dimenzation
of the vanadium atoms, the corresponding band for the $M_1$ phase is split into
bonding and antibonding subbands and, which is more important, the filling of
the $xy$ orbital becomes maximum between the $t_{2g}$ orbitals. The nonmagnetic
LDA calculation does not reproduce the insulating nature of the energy spectrum
for the $M_1$ phase of VO$_2$. Yet, it is clear that the allowance for
correlations in calculations will not change this key result, i.e., the
preferential filling of the orbitals of the $xy$ symmetry.

The conclusion concerning filling of the $3d^1$ $xy$ states of the $M_1$ phase
is supported by two factors. First, the distortions of the oxygen octahedra
around the vanadium atom occur in such a way that the maximum V-O distances are
observed in the $xy$ plane in the local system of coordinates (see table),
which, in turn, lowers the energy of the $xy$ orbital. Second, we performed the
LSDA+U calculations for the $M_1$ structure to determine which of the orbitals
will be filled by the localized electron\footnote{In calculating the $M_1$ and
$M_2$ phases of VO$_2$, it is incorrect to substantiate conclusions by
comparing the centers of gravity of orbitals and the dispersions of the
corresponding bands obtained in nonmagnetic LDA calculations in view of the
lower symmetry of these phases and substantial hybridization of the states
under investigation.}. Note that the LSDA+U calculation predicts for the $M_1$
phase the antiferromagnetic ground state with a gap equal to 0.55~eV and the
spin magnetic moment of the $d$ shell of vanadium $\mu$=0.81~$\mu_B$ instead of
the nonmagnetic insulating state with a gap of 0.2-0.7~eV observed
experimentally [9, 27]. Quantum corrections, probably, could suppress the
long-range magnetic order obtained in the LSDA+U calculation, and the
calculated value of the gap is within the limits of the experimental one. Yet,
turning to the discussion of the orbital nature of the localized $d$ electron,
one can see from Fig. 5 that the $xy$ orbital became localized. Both factors
permit us to conclude that upon transition of VO$_2$ from the $R$ to the $M_1$
phase, along with a change in the ratio of the distances from the vanadium atom
to the in-plane oxygen atom and to the apical oxygen atom, an important
transformation of orbital symmetry of the localized $3d^1$ electron of the
V$^{4+}$ ion occurs, namely, from the $yz-zx$ to the $xy$. 

We believe that the Peierls mechanism is not responsible for the formation of
the insulating state of the $M_1$ phase. Even the LDA results show that the
most filled orbitals in both cases are different: the $xy$ orbital for the
$M_1$ phase and the $yz-zx$ orbital for the $R$ phase. It is known (for
example, from the theoretical investigations of CrO$_2$~\cite{Korotin98} which
also has a crystal structure of the rutile type) that the $yz-zx$ band has
substantial dispersion and is strongly hybridized with the O$2p$ states,
forming the $\pi$ bond with them. Therefore, it is weakly localized even in the
presence of strong correlations and the energy spectrum may remain metallic in
the case of the preferential filling of this band. At the same time, the $xy$
band is narrower than the $yz-zx$ one and more sensitive to correlations. This
results in localization of electrons on the $xy$ orbital, its complete filling,
and the appearance of a gap in the energy spectrum. It is what we obtained in
our LSDA+U calculations. Thus, we can conclude that it is the Mott-Hubbard
mechanism that is responsible for the formation of the energy gap in VO$_2$. The
Peierls's mechanism can be used only for explanation of the insulating state of
the $M_1$ phase, but it is completely unacceptable for the $M_2$ phase, in
which both dimerized and undimerized chains of vanadium atoms exist
simultaneously. Taking into account which orbitals are filled by the $d$
electron in each particular case and which are the properties of these
orbitals, the Mott-Hubbard mechanism can be applied for the description of both
metallic and insulating states of all crystal phases of VO$_2$. 

The calculations of the electronic structure of the $M_2$ phase (space group
$C2/m$) fully confirm our assumption about the mechanism of formation of energy
bands in VO$_2$. Both dimerized chains of vanadium atoms (similar to the chains
in the $M_1$ phase) and the undimerized ones (as in the $R$ phase) exist in the
crystal structure of the $M_2$ phase. As in the $M_1$ phase, but contrary to
the $R$ phase, the V-O distances to the apical oxygen atoms in the oxygen
octahedra which surround the vanadium atoms in both types of chains are smaller
than to the in-plane oxygens. The Euler angles for the passage to the local
system of coordinates for the vanadium atoms in the dimerized and undimerized
chains (Fig. 6) are given by the sets of $(-\pi; -\frac \pi 4; \frac \pi 4)$
and $(0; -\frac \pi 4; \frac \pi 4)$. 

The "nonmagnetic" calculations in the local-density approximation give the
metallic ground state, which contradicts the experiment. Owing to a substantial
distortion of the crystal structure and, correspondingly, complicated
hybridization of the $d$ states of vanadium with each other, it is impossible
here to consistently separate the partial densities of the $t_{2g}$ states in
the cubic harmonics, which we had done, for example, for the $R$ phase. The
LSDA+U calculation is the sole possibility of determining which localized
$t_{2g}$ orbital will be filled. 

\begin{figure}[t]
\centerline{\includegraphics [clip=true,width=75mm]{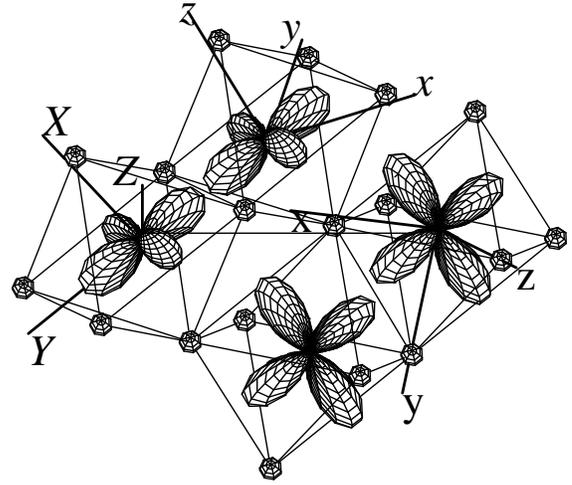} } 
\caption{Ordering of the occupied $xy$ orbitals in the dimerized (upper left
part of the figure) and the undimerized (lower right part of the figure) chains
of the vanadium atoms obtained by the LSDA+U calculation for the $M_2$ phase of
VO$_2$. The axes of the crystallographic system of coordinates are marked by
capital letters. The axes of the local systems of coordinates are shown by
small letters.} 
\label{M2.orb} 
\end{figure}

The results of such a calculation are shown in Fig.~\ref{M2.orb}. They
demonstrate the occupied orbitals of vanadium that in the $M_2$ phase are the
$xy$ orbitals both in the dimerized and undimerized chains. According to
calculations, in the case of antiferromagnetic ordering the spin magnetic
moment of the $d$ shell of vanadium that belongs to the undimerized chain is
equal to $\mu$ = 0.89~$\mu_B$ and the arising energy gap is 0.93~eV. We also
calculated the exchange-correlation parameter $J$ of the Heisenberg model for
the vanadium atoms in the undimerized chain (the scheme of calculation within
the LSDA+U method is described in \cite{lda+u_modern}. Its value was found to
be 205~K and the sign corresponds to the antiferromagnetic exchange. The value
obtained is close to the result of \cite{Pouget74} where, from an analysis of
nuclear magnetic resonance experiments, it is shown that the V$^{4+}$ ions in
the undimerized chains behave as Heisenberg chains with spin $S = 1/2$ and
antiferromagnetic exchange J$\approx$300~K.

Thus, the principal difference between the rutile $R$ phase and the monoclinic
$M_1$ and $M_2$ phases is which of the localized $t_{2g}$ orbitals is filled by
the sole $3d$ electron of the V$^{4+}$ ion: $yz-zx$ or $xy$. The choice of the
filled orbital is determined by the local oxygen surroundings of the vanadium
atoms. Since these bands have different properties relative to localization (or
dispersion), the correlation effects variously influence the energy spectrum
(metal or insulator) depending on which band is filled. It is correlations (the
Mott-Hubbard mechanism) that are responsible for the insulating properties of
the $M_1$ and $M_2$ phases of VO$_2$. Correlations are also present in the
metallic $R$ phase. Owing to the localization of the $d$ electron in the
$yz-zx$ orbital, they do not lead to the appearance of a gap in the band
spectrum.

This work was supported by the Russian Foundation for Basic Research, project
no. 01-02-17063, and by NWO (grant no. 047-008-012).

\pagestyle{empty}

\end{document}